\documentclass[10pt,conference]{IEEEtran}
\IEEEoverridecommandlockouts

\usepackage{blindtext}
\usepackage{cite}
\usepackage{amsmath,amssymb,amsfonts}
\usepackage{algorithmic}
\usepackage{textcomp}
\usepackage{xcolor}
\usepackage{amsmath,amssymb,amsfonts}
\usepackage{graphicx}
\usepackage{subfigure}
\usepackage{enumitem}
\usepackage[nolist,nohyperlinks]{acronym}

\usepackage{booktabs} 

\usepackage[font=small]{caption}

\usepackage{lipsum}

\begin{document}

\title{A Novel Metric for mMIMO Base Station Association for Aerial Highway Systems}

\author{\IEEEauthorblockN{Matteo Bernabè\IEEEauthorrefmark{1}\IEEEauthorrefmark{4}, 
David López Pérez\IEEEauthorrefmark{2},
Nicola Piovesan\IEEEauthorrefmark{1},
Giovanni Geraci\IEEEauthorrefmark{3}, and
David Gesbert\IEEEauthorrefmark{4}}
\\ \vspace{-0.3cm}
\normalsize\IEEEauthorblockA{\emph{\IEEEauthorrefmark{1}Huawei Technologies, France \enspace \IEEEauthorrefmark{2}Univ. Politècnica de València, Spain \enspace \IEEEauthorrefmark{3}Univ. Pompeu Fabra, Spain \enspace \IEEEauthorrefmark{4}EURECOM, France}
}
\thanks{This research was supported  by the Generalitat Valenciana, Spain, through the  CIDEGENT PlaGenT,  Grant CIDEXG/2022/17,  Project iTENTE, and by the Spanish State Research Agency through grant PID2021-123999OB-I00 and the ``Ram\'{o}n y Cajal'' program.} 

}

\bstctlcite{IEEEexample:BSTcontrol}

\maketitle

\begin{abstract}\label{abstract}
In this article, 
we introduce a new metric for driving the serving cell selection process of a swarm of cellular connected unmanned aerial vehicles (CCUAVs) located on aerial highways when served by
a massive multiple input multiple output (mMIMO) terrestrial network.
Selecting the optimal serving cell from several suitable candidates is not straightforward.
By solely relying on the traditional cell selection metric, 
based on reference signal received power (RSRP),
it is possible to result in a scenario in which the serving cell can not multiplex an appropriate number of CCUAVs due to the high correlation in the line of sight (LoS) channels.
To overcome such issue, 
in this work, 
we introduce a new cell selection metric to capture not only signal strength, 
but also spatial multiplexing capabilities.
The proposed metric highly depends on the relative position between the aerial highways and the antennas of the base station.
The numerical analysis indicates that the integration of the proposed new metric allows to have a better signal to interference plus noise ratio (SINR) performance on the aerial highways,
resulting in a more reliable cellular connection for CCUAVs.
\end{abstract}

\section{Introduction}\label{sec:Introduction}

Remote piloted drones, also known as \acp{UAV}, have become increasingly important in recent years,
having already had a major impact on different applications, 
such as surveillance and security, precision agriculture, and parcel delivery~\cite{UAV_Communications_for_5G_and_Beyond,9768113}.
In May 2021, 
Morgan Stanley predicted that, 
by 2050,
the entire \ac{UAM} market, 
including air taxis, delivery, and patrol drones, 
could reach a value up to \$19 trillion, 
accounting for 10 to 11\%  of the projected United States global gross domestic product (GDP)~\cite{MorganStanley}.
In addition, 
it is expected that the intrinsic flexibility of \acp{UAV} will enable new disruptive industries and markets that are currently beyond our imagination.

The use of \acp{UAV} in communication networks can be categorized into two main categories:
\emph{i)} UAV-aided networks, 
where \acp{UAV} act as flying base stations, or relays,
and \emph{ii)} \acp{CCUAV}, 
where \acp{UAV} connect to the network as flying \ac{UE}. 
In both categories, 
supporting \acp{UAV} with a reliable connection is essential for safe and effective operation.
Cellular network connectivity provides a promising solution to this challenge,
allowing \acp{UAV} to communicate with ground control stations over long distances \ac{BVLoS}.

Given a fourth and/or fifth generation (4G/5G) cellular network,
to provide a minimum \ac{QoS} with reliability guarantees, 
e.g., 100~kbps rate and 50~ms latency at 3 nines of reliability for the command and control (C\&C) channel of a \ac{CCUAV}~\cite{3GPP36777},
most of the research has focused on the optimization of the trajectory of the \ac{CCUAV}~\cite{Cherif_disconnecivityAware,esrafilian20203d, Challita_PathPlanningIntereferenceAwareRL}.

Despite the importance of UAV trajectory optimization,
to support the significant growth and expansion of \ac{UAV} applications,
authorities and industries are working towards the creation of an organised system of \ac{UAV} highways in the sky to facilitate operation management and ensure reliable connectivity on predetermined aerial routes planned according to government and/or business criteria~\cite{PilotsHandbook_FAA, 3DAerialHighway_Magazine}.
Thus, optimizing 4G, 5G networks to support a minimum \ac{QoS} with reliability guarantees over a limited segregated airspace may be a more feasible and practical approach than route optimization over a given network.

The research community has begun to adopt such a complementary approach. 
However, only a few pioneering works exist in the literature. 
In~\cite{karimi2023analysis},
the authors carried out a mathematical analysis of the \ac{RSS} perceived by \acp{CCUAV} flying on aerial corridors, 
while being served by a ground cellular network.
In~\cite{chowdhury2021ensuring}, 
the authors explored the deployment of a new set of base stations with uptilted antennas to specifically serve aerial highways.
They also propose an \ac{eICIC} technique to mitigate interference to/from the aerial corridors. 
Similarly, in~\cite{PlacementofmmWaveAntenna}, 
the authors proposed a framework to optimize the deployment of uptilted \ac{mmWave} access points to serve \acp{CCUAV} on aerial highways.
In our previous work~\cite{10001469},
instead of deploying new base stations for \acp{CCUAV},
we developed a stochastic ADAM-based optimization algorithm to fine-tune the downtilt of an existing 4G macrocellular network to maximize the \ac{CCUAV} and ground UE rates, 
while providing a minimum SINR performance on the predefined aerial highways.

In recent years, 
various other solutions based on, e.g., null steering, \ac{D2D} communications, have been investigated to ensure a \ac{CCUAV} reliable connectivity provided a cellular network~\cite{8528463, 10008629, 10001193}.
However, none of the mentioned frameworks have investigated the importance of \ac{CCUAV} cell association to the ground macrocellular network. 
Given that multiple \acp{CCUAV} will be closely located over the aerial highway, 
selecting the serving cell that provides the largest \ac{RSRP} may be suboptimal as it may not allow to efficiently exploit \ac{mMIMO} multiplexing capabilities. 
In this paper, 
we investigate a new metric to drive the \ac{CCUAV} cell association process to a \ac{mMIMO} 5G network.
Our proposed methodology uses planning information collected across the aerial highway to extrapolate the \ac{mMIMO} multiplexing capabilities of a cell over a given route.
This information, 
in addition to the \ac{RSRP},
is then incorporated into the \ac{CCUAV} cell selection logic to select the best server among the suitable set of candidates.

The rest of the paper is organized as follows.
In Section~\ref{sec:SystemModel}, 
we introduce the adopted system model. 
In Section~\ref{sec:CellAssociationandMetrics}, 
we define the investigated cell selection metric for \acp{CCUAV}. 
In Section~\ref{sec:Results}, 
we discuss our experiments and results, and finally, 
in Section~\ref{sec:Conclusion}, the conclusions are drawn.

\section{System Model}\label{sec:SystemModel}

To illustrate the main concept behind the proposed metric, 
we consider a sub-6GHz downlink scenario comprised of $N_\mathrm{MS}=3$ macrocellular sectors covering an area of $A$~km$^2$.
Each sector operates at carrier frequency $f_c$, 
and is equipped with a planar \ac{mMIMO} antenna panel with $M$ active antennas, 
located at a height of $h_s=25$\,m.
To provide more detail, 
the \ac{mMIMO} antenna panel is composed of $M_h$ antennas on the horizontal axis and $M_v$ on the vertical one, 
i.e., $M=M_h \times M_v$. 
The distance between each antenna element is $\lambda_p / 2$, 
where $\lambda_p $ is the design wavelength. 
The \ac{mMIMO} antenna panel is directed toward the center of the mentioned covered area.
%
The total transmit power of each sector and the transmit power allocated to each \ac{PRB} by each sector are equal to $P_{T_x}^\mathrm{Tot}$ and $P_{T_x}$, respectively. 
Note that in this paper, 
we consider that the total transmit power $P_{T_x}^\mathrm{Tot}$ of the sector is equally divided among the $N_\mathrm{PRB}$ \acp{PRB} managed by the sector, 
i.e., $P_{T_x} = P_{T_x}^\mathrm{Tot} / N_\mathrm{PRB}$.

A set of $R$ highways of length $L_r$ are deployed at the center of the scenario at an altitude $h_a$. 
Such highways share the same center point, 
and are symmetrically rotated from each other by an angle $\Delta_\phi$.
Each highway is defined by $N_w$ equidistant aerial waypoints (i.e., reference points),
with an inter-waypoint distance $d_w=1$~m.

\begin{figure}[!t]
    \centering
    \includegraphics[width=0.48\textwidth]{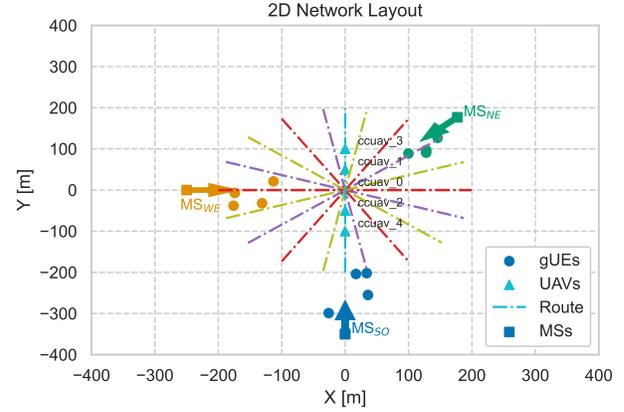}
    \caption{2D Network Layout with 3 MS with $N_g$ gUEs each, $R$ routes and $N_\mathrm{ccuav}$ CCUAVs positioned on the horizontal lane.}
    \label{fig:2DNetworkLayout}
\vspace{-1.5em}
\end{figure}

$N_\mathrm{ccuav}$ single-antenna \acp{CCUAV} are then deployed on the aerial waypoints of each highway.  
Following the recommendations in~\cite{Vinogradov_reducingSafeUAVDistance},
the chosen inter-\ac{CCUAV} distance is $d_\mathrm{ccuav}=50$~m.
In addition to the \acp{CCUAV}, 
$N_g$ single-antenna \ac{gUE} are also randomly deployed within the coverage area of each sector.
Figure~\ref{fig:2DNetworkLayout} depicts the network layout and the deployed aerial highways with \acp{CCUAV} positioned on one of them.

For the sake of clarity,
let us denote by $\mathbb{B}$ the set of sectors, 
with $\mathrm{Card}\left\{\mathbb{B}\right\} = N_\mathrm{MS}$,
and by $\mathbb{D}$, $\mathbb{G}$ and $\mathbb{U}$ the sets of \acp{CCUAV}, \acp{gUE} and all \acp{UE} in the network, respectively,
such that $\mathrm{Card}\left\{\mathbb{D}\right\} = N_\mathrm{ccuav}$, $\mathrm{Card}\left\{\mathbb{G}\right\} = N_g$ and $\mathrm{Card}\left\{\mathbb{U}\right\} = N_\mathrm{ccuav} + N_g$, respectively.

For the sake of simplicity, 
let us assume that a sector multiplexes all its connected \acp{gUE} and \acp{CCUAV} across its bandwidth ---all its \acp{PRB}.


\subsection{Channel Model}

The large-scale channel characteristics, 
\ac{LoS} probability, path loss, and shadow fading, 
for \acp{gUE} and \acp{CCUAV} are modelled according to the \ac{3GPP} urban macro scenarios in~\cite{3GPP38901} and~\cite{3GPP36777}, respectively.
We also integrate 2D spatial correlation features for the stochastic lognormal shadowing between \acp{UE} and sectors. 
To compute such realizations, 
we adopt the sum of sinusoids approach presented in \cite{ShadowCorrelation_Xiaodong}.

The complex channel, $\textbf{h}_{u, b} \in \mathbb{C}^{1XM}$, between a \ac{UE} $u \in \mathbb{U}$ and the $M$ antennas of the \ac{mMIMO} panel of sector $b \in \mathbb{B}$ is modeled as a Rician random variable, i.e.,
\begin{equation}\label{eq:General_RicianChannel}
\textbf{h}_{u, b} = \sqrt{\frac{K}{1+K}}\; \textbf{h}_{u, b}^{\mathrm{LOS}} + \sqrt{\frac{1}{1+K}}\; \textbf{h}_{u, b}^{\mathrm{NLOS}},
\end{equation}
with 
\begin{equation}
    \textbf{h}_{u, b}^\mathrm{LOS} = e^{j \frac{2 \pi}{\lambda_c} \textbf{d}_{u,b}},
\end{equation}
and
\begin{equation}
     \textbf{h}_{u, b}^\mathrm{NLOS} \sim  \mathbb{CN}^M(0,1) \,\,,
\end{equation}
where $\textbf{d}_{u,b}= \left[d_{u,b}^0, \ldots, d_{u,b}^M \right]$ is the distance between a \ac{UE} $u$ and an antenna element $m$ of the \ac{mMIMO} panel of sector $b$,
and $K$ is the $K$-Rician factor, 
whose value is specified in~\cite{3GPP36777}.

\subsection{Signal quality model}

The signal $y_{u}$ received at \ac{UE} $u$ from its serving sector $s$ is formulated as
\begin{gather}\label{eq:received signal}
    y_u = \sqrt{ \beta_{u,s} }  \,  
    \textbf{h}^H_{u,s}  \textbf{w}_{u,s} + \\  \nonumber
    \sqrt{\beta_{u,s}}  
    \sum_{p \in \mathbb{U}_s \setminus u} \textbf{h}^H_{u,s}  \textbf{w}_{p,s} + \\  \nonumber
    \sum_{b \in \mathbb{B} \setminus s } \sqrt{\beta_{u,b}}  
    \sum_{i \in \mathbb{U}_b} \textbf{h}^H_{u,b} \textbf{w}_{i,b} + n_u \,,
\end{gather}
with 
\begin{equation}\label{eq:LargeScale_gain}
    \beta_{u,b} = P_{T_x} \, G_{u,b} \, \rho_{u,b} \, \tau_{u,b} \,,
\end{equation}
where $\mathbb{U}_b \subset \mathbb{U}$ is the set of \acp{UE} connected to sector $b$, 
$s \in \mathbb{B}$ is the serving sector of \ac{UE} $u$, 
$G_{u,b}$, $\rho_{u,b}$ and $\tau_{u,b}$ are the antenna, path and shadow fading gains between \ac{UE} $u$ and sector $b$, respectively,
$\textbf{w}_{u,b} \in \mathbb{C}^{M \times 1}$ is the precoding vector devised to serve \ac{UE} $u$ at sector $b$,
and $n_u$ is the thermal noise.

At a given sector $b$, 
and for a given channel matrix
$\textbf{H}_b = \left[  \textbf{h}_{1,b}, \textbf{h}_{2,b}, \ldots, \textbf{h}_{N_b^\mathrm{in},b}  \right]^T$,
the precoding matrix $\textbf{W}_b = \left[  \textbf{w}_{1,b}, \textbf{w}_{2,b}, \ldots, \textbf{h}_{N_b^\mathrm{in},b}  \right]$
is derived using \ac{ZF} as~\cite{8528463}
\begin{equation}\label{eq:ZeroForcing}
    \textbf{W}_b = \hat{\textbf{H}}_b^H \left( \hat{\textbf{H}}_b \; \hat{\textbf{H}}_b^H \right)^{-1} \textbf{D}_b^{-1/2},
\end{equation}
where $N_b^\mathrm{in}$ is the cardinality of set $\mathbb{U}_b$ 
(i.e., the number of \acp{UE} served by sector $b$),
$\hat{\textbf{H}}_b$ is the estimated channel matrix,
and $\textbf{D}_b^{-1/2}$ is the diagonal normalization matrix defined to satisfy the transmit power constraints, 
with an equal transmit power allocation for each \ac{UE} in this case.
%
%
%

In this work, 
we assume that the set of pilot signals used for channel estimation at each sector is orthogonal with respect to those used at the other 2 sectors. 
This results in perfect \ac{CSI}, 
and thus the estimated and the real channel matrices coincide perfectly, 
i.e., $\hat{\textbf{H}}_b = \textbf{H}_b$.

Finally, 
the resulting \ac{SINR} at \ac{UE} $u$ when associated to sector $s$ is defined as
\begin{equation}
    \gamma_u = \frac{ P_u  }  
    {I_u + N_u} = \frac{ P_u  }  
    {\left( I_u^{\mathrm{intra}} + I_u^{\mathrm{inter}}\right) + N_u},
\end{equation}
with
\begin{equation}\label{eq:UsefulPowerFromula}
    P_u = \beta_{u,s} 
    \left|\textbf{h}^H_{u,s}  \textbf{w}_{u,s} \right|^2 \,,
\end{equation}
and
\begin{gather} \label{eq:InterferenceFormula}
    I_u = I_u^{\mathrm{intra}} + I_u^{\mathrm{inter}} = \\ \nonumber
    \sqrt{\beta_{u,s}}  
    \sum_{p \in \mathbb{U}_s \setminus u} \left|\textbf{h}^H_{u,s}  \textbf{w}_{p,s}\right|^2 +
    \sum_{b \in \mathbb{B} \setminus s } \sqrt{\beta_{u,b}}  
    \sum_{i \in \mathbb{U}_b} \left| \textbf{h}^H_{u,b} \textbf{w}_{i,b}\right|^2 \,,
\end{gather}
where $P_u$ is the useful received power at \ac{UE} $u$, 
$I_u$ is the total interference composed of the intra-cell interference $I_u^{\mathrm{intra}}$ and the inter-cell interference $I_u^{\mathrm{inter}}$, 
whereas $N_u$ is the thermal noise power.
Note that, 
when using \ac{ZF} and under perfect \ac{CSI},
 $I_u^{\mathrm{intra}}=0$ if $N_b^\mathrm{in} \leq M$.

\section{Serving Sector Association} \label{sec:CellAssociationandMetrics}

In traditional cellular networks, 
the serving sector of each \ac{UE} is typically determined using metrics related to \ac{RSRP}. 
However, when considering \acp{CCUAV} operating at altitudes above 100\,m, 
most of them are likely to be in \ac{LoS} with many sectors, 
resulting in a high probability of measuring a comparable \ac{RSRP} from all of them~\cite{3GPP36777}. 
This leads to two major drawbacks: 
\emph{i)} frequent handovers and ping-pong effects, and
\emph{ii)} poor experienced \acp{SINR}, 
driven by the high inter-cell interference. 
Importantly,
it should be noted that when employing \ac{ZF} precoding with closely-located \acp{CCUAV},
the high correlation between the complex channels of nearby \acp{CCUAV} can also lead to a noise-enhancement problem, 
arising from the increased values in the normalization matrix $\textbf{D}_b^{-1/2}$,
which decreases the useful received power $P_u$.
This further affects \acp{CCUAV} performance.

In addition to the advantage of using aerial highways to manage aerial operations, 
knowing the a priori flight route allows extrapolating useful information on the \acp{AoA} of \acp{CCUAV},
which can help to alleviate some of the above challenges.
In the following, 
we concentrate on a solution for the following drawback.
In conventional cellular networks, 
\acp{CCUAV} typically connect to the sector that provides the strongest \ac{RSRP}.
However, the independent design of the aerial trajectory with respect to the existing cellular network may result in scenarios where the strongest sector is unable to resolve the \ac{mMIMO} spatially multiplexed communications to/from \acp{CCUAV} flying on such route and seen at approximately the same \ac{AoA}.
Such channel correlation would result in the mentioned noise-enhancement problem, 
and consequently poor overall \acp{CCUAV} \ac{SINR} performance.
An example of such scenario would be that where the aerial route is perpendicular to the \ac{mMIMO} antenna panel of the strongest sector. 
In those cases, 
it may be beneficial to associate to a reasonably weaker sector but with better multiplexing capabilities. 
This is a trade-off exacerbated by the nature of aerial highways,
which has never been investigated in the \acp{CCUAV} literature. 
To address the aforementioned issues,
this study proposes a novel indicator that relies on the exploitation of predefined aerial highways design information to assign a multiplexing capability score to each sector, 
which can be used to drive a smarter cell selection process for \acp{CCUAV}.

\subsection{Eigenscore-based Indicator }\label{subsec:eigenscore}

In this section, 
we introduce a new cell selection indicator for enhancing the cell selection ---and thus the performance--- of \acp{CCUAV} flying on aerial highways.  
For the sake of argument, 
let us consider a route $r$ with its $N_w$ equidistant aerial waypoints, 
as mentioned earlier.

In a planing stage, 
a complex channel vector $\textbf{h}_{w, b}$ between each waypoint $w$ and sector $b$ can be calculated by performing a set of measurements, 
before starting to operate the aerial route, 
and/or using eq.~\eqref{eq:General_RicianChannel} in this case.
Then, the complex channel vectors $\textbf{h}_{w, b}$ collected across all waypoints can be used to create the complex channel matrix $\textbf{H}_{r,b} \in \mathbb{C}^{N_w \times M}$ associated with route $r$ and sector $b$.
Once the complex channel matrix $\textbf{H}_{r,b}$ is derived for each route $r$ and sector $b$,
the related set of eigenvalues $\boldsymbol{\Lambda}^{\mathrm{eig}}_{r,b}$ can be calculated using,
e.g., \ac{SVD}, 
and scaled in the range $\left[0,1\right]$ to proportionally identify the most relevant ones as follows
\begin{equation}
    \boldsymbol{\bar{\Lambda}}_{r,b}^{\mathrm{eig}} = \frac{\boldsymbol{\Lambda}^{\mathrm{eig}}}
    {\sum_{ \lambda_i \in \boldsymbol{\Lambda}^{\mathrm{eig}}_{r,b}} \left| \lambda_i \right|^2} \,.
\end{equation}
With this, 
we can define an eigenscore ${\rm ES}_{r,b}$ for each route $r$ and sector $b$ as the number of eigenvalues greater than a threshold $\lambda_{\mathrm{Th}} \in \left[ 0,1\right]$,
i.e.,
\begin{equation}\label{eq:EigenScore_definition}
    {\rm ES}_{r,b} = \mathrm{Card}\left\{\mathbb{K}_{r,b} \right\},  \mathrm{with} \, \, \mathbb{K}_{r,b} =
    \left\{ k \in   \boldsymbol{\bar{\Lambda}}_{r,b}^{\mathrm{eig}} \,
    | \, k \geq \lambda_{\mathrm{Th}}  \right\} \,.
\end{equation} 

\begin{figure}[!t]
    \centering
    \subfigure[Eigenvalues south ($\mathrm{MS_\mathrm{SO}}$).]{
    \includegraphics[width=0.2275\textwidth]{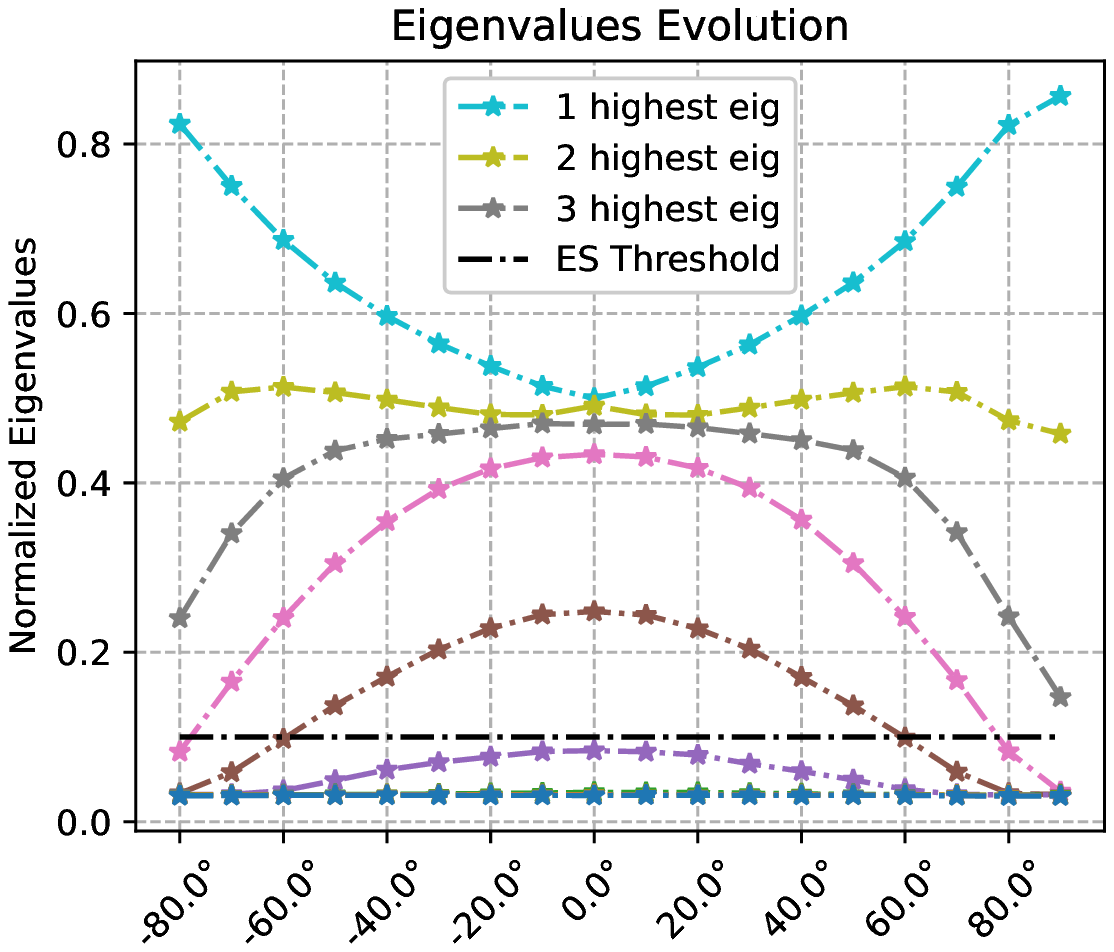}
    \label{subfig:Eigvalues_bs0}
    } 
    \subfigure[Eigenscore south ($\mathrm{MS_\mathrm{SO}}$).]{
    \includegraphics[width=0.2275\textwidth]{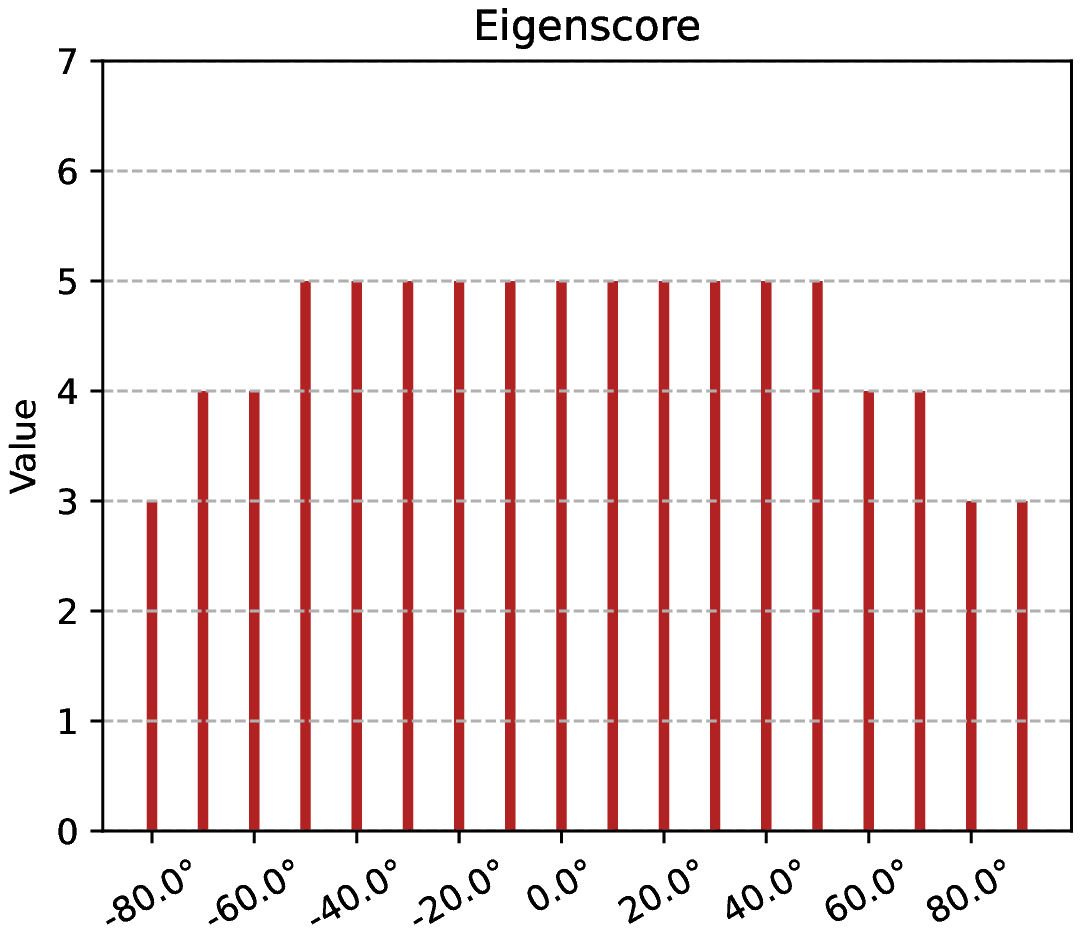}
    \label{subfig:Eigscore_bs0}
    }
    \subfigure[Eigenvalues west ($\mathrm{MS_\mathrm{WE}}$).]{
    \includegraphics[width=0.2275\textwidth]{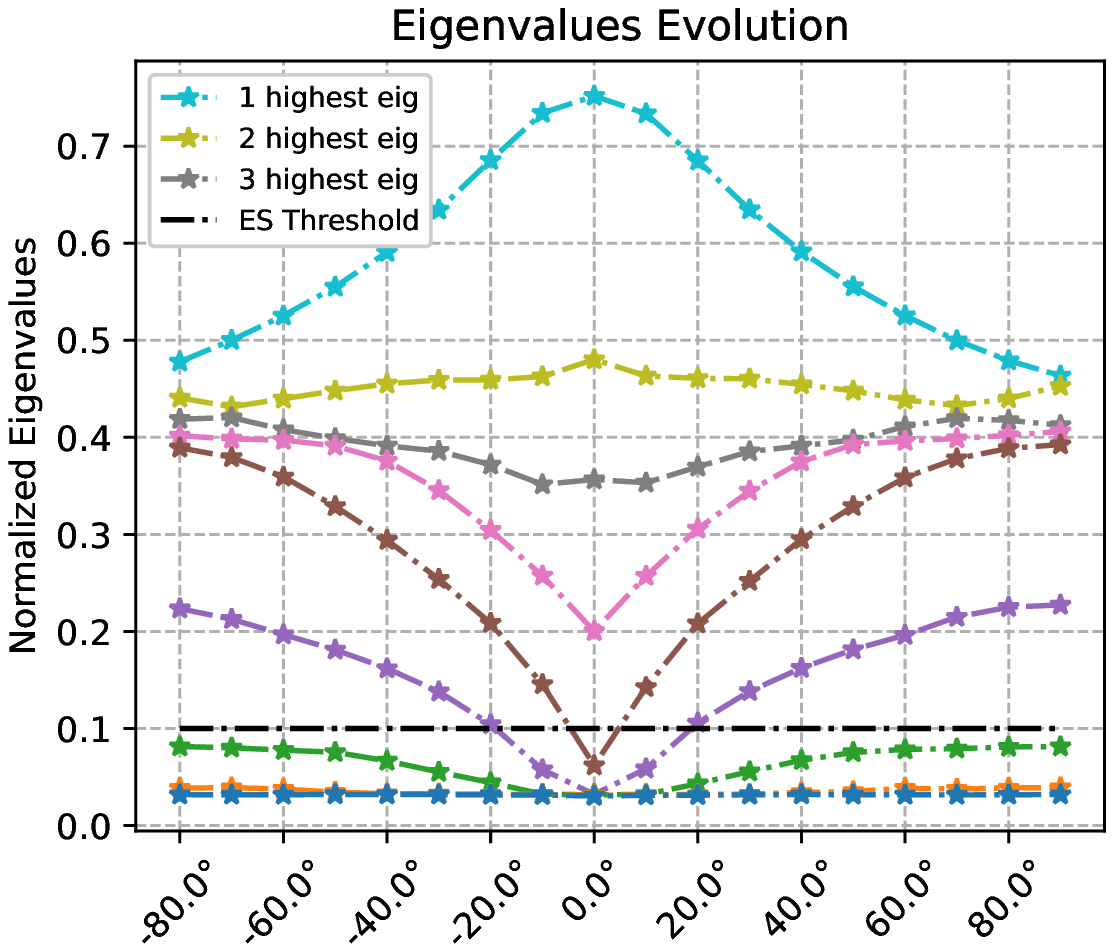}
    \label{subfig:Eigvalues_bs1}
    }
    \subfigure[Eigenscore west ($\mathrm{MS_\mathrm{WE}}$).]{
    \includegraphics[width=0.2275\textwidth]{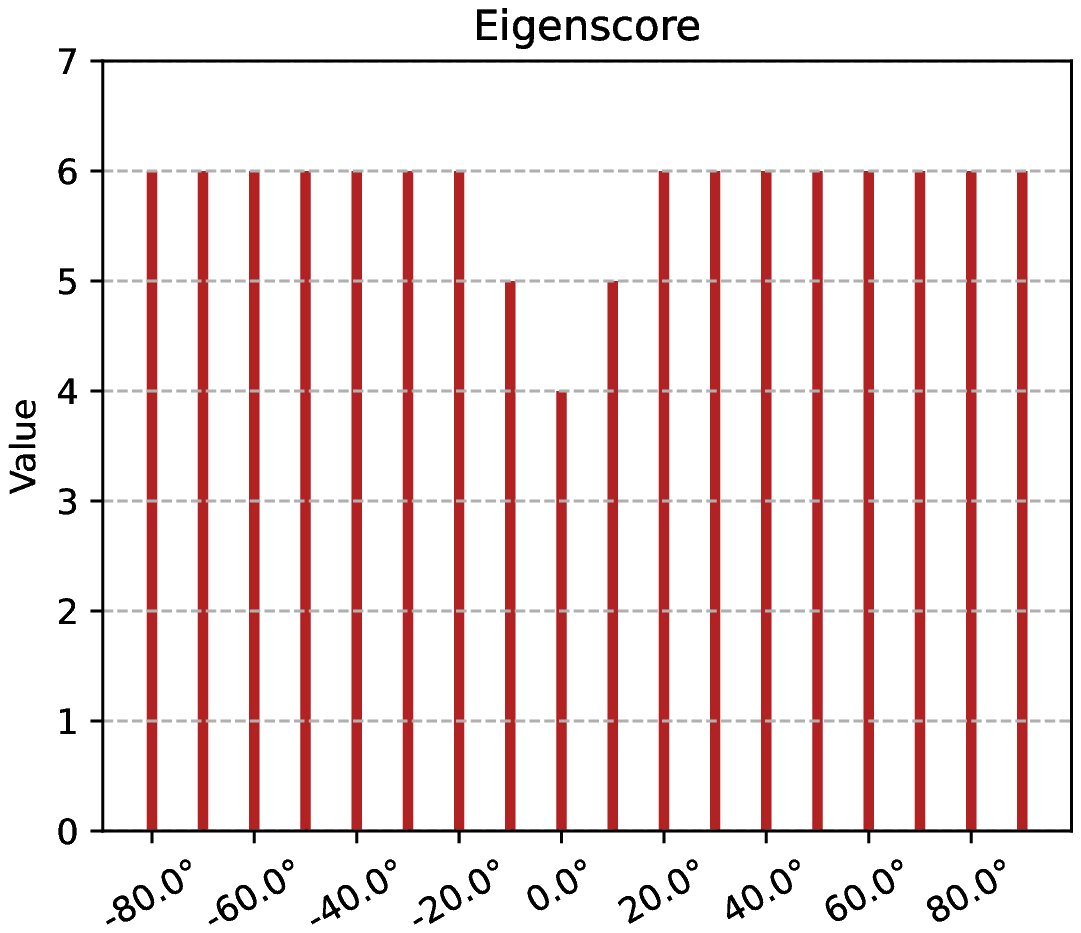}
    \label{subfig:Eigscore_bs1}
    }
    \subfigure[Eigenvalues north-east ($\mathrm{MS_\mathrm{NE}}$).]{
    \includegraphics[width=0.2275\textwidth]{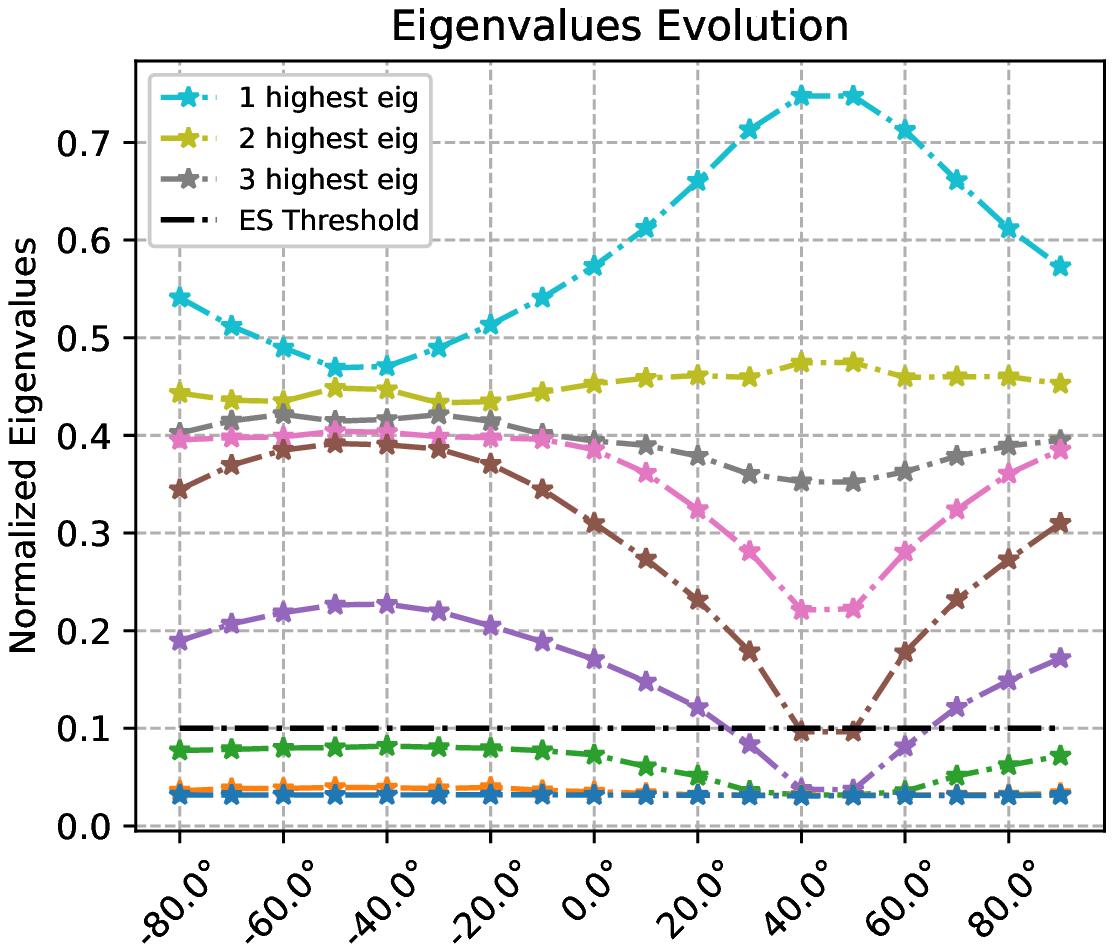}
    \label{subfig:Eigvalues_bs2}
    }
    \subfigure[Eigenscore north-east ($\mathrm{MS_\mathrm{NE}}$).]{
    \includegraphics[width=0.2275\textwidth]{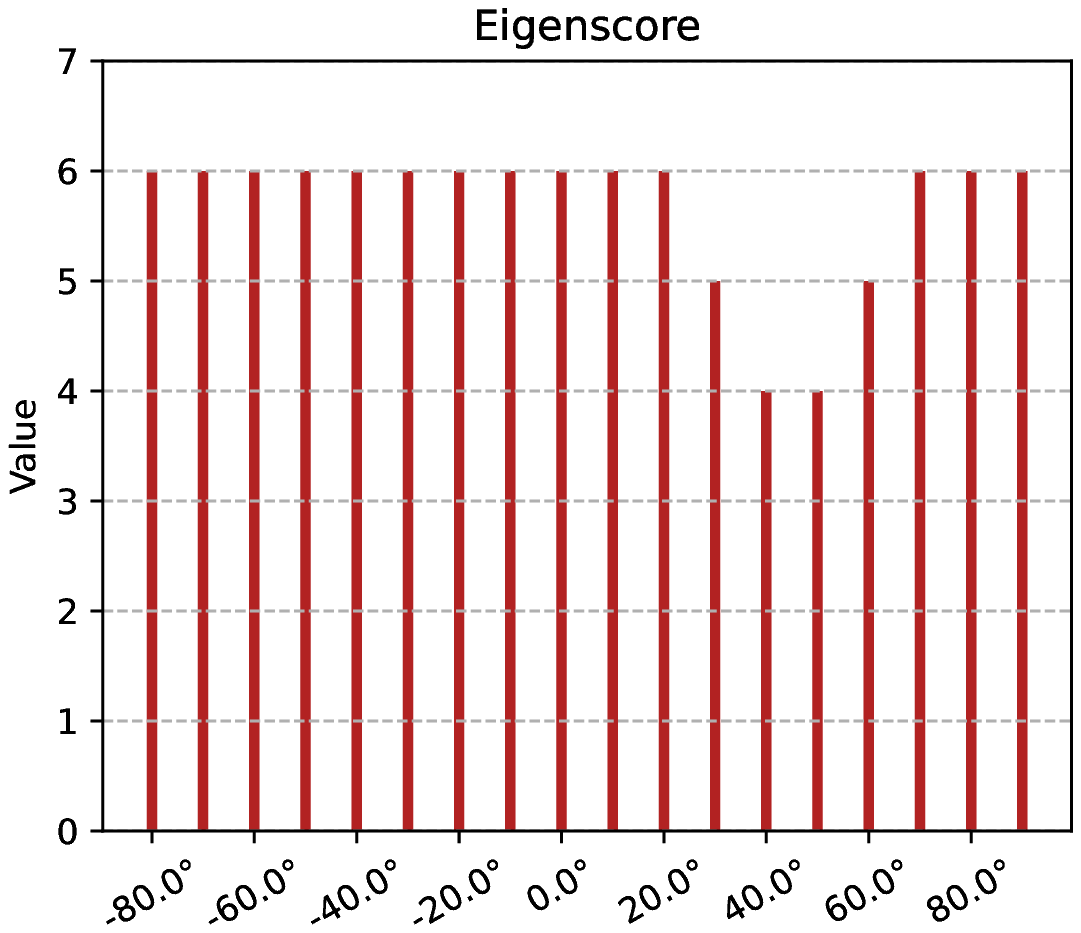}
    \label{subfig:Eigscore_bs2}
    }
\caption{Eigenvalues and Eigenscore computed in each sector for different rotation or aerial highways. With threshold $\lambda_{\mathrm{Th}}=0.10$\,. \label{fig:EigenvaluesEigenscore_evolution}}
\vspace{-0.965cm}
\end{figure}

It is worth highlighting that the defined eigenscore ${\rm ES}_{r,b}$ is highly dependent on the geometry of the problem, 
e.g., the angle of the aerial highway with respect to the \ac{mMIMO} antenna panel. 
For instance, 
a route $r$ precisely aligned with the normal direction of the \ac{mMIMO} antenna panel of sector $b$ and located at the same altitude, 
will yield the eigenscore ${\rm ES}_{r,b}=1$.
Conversely, 
a route $r$ parallel to the \ac{mMIMO} antenna panel of sector $b$ is likely to exhibit a higher eigenscore, 
i.e., ${\rm ES}_{r,b}>1$
as the incoming signals are perceived with distinct \acp{AoA}.
Plainly speaking, 
this eigenscore allows to assess the degrees of freedom that sector $b$ has on route $r$.
To corroborate such statements, 
Figure~\ref{fig:EigenvaluesEigenscore_evolution} illustrates, 
for each sector $b$ in our network, 
the eigenvalues and eigenscores when assessing various routes with different orientations at an altitude of 100\,m.

In this work,
we are interested in considering both the signal strength and spatial diversity features of the complex channel to drive the cell selection process,
with minimal changes to the state-of-the-art process.
In the following, 
we show how this new indicator blends with the traditional \ac{RSRP}-based one.

%

\subsection{Enhanced Cell Selection Metric}

We now exploit our proposed eigenscore ${\rm ES}_{r,b}$ in eq.~\eqref{eq:EigenScore_definition} to formulate two new metrics to drive the \acp{CCUAV} cell selection process. 
Let us assume that,
in the operational stage, 
each sector $b$ broadcasts for each route $r$ the earlier calculated eigenscore $\mathrm{ES_{r,b}}$ during the planning stage in its broadcast channel. 
In addition, 
let us assume that all \acp{CCUAV} are aware of a maximum $\mathrm{RSRP}_\mathrm{}^\mathrm{max}$ and a minimum $\mathrm{RSRP}_\mathrm{}^\mathrm{min}$ \ac{RSRP} value for normalization purposes. 
The sector $b$ with the resulting larger metric will be selected as server. 

With this in mind,
and dropping the route index $r$ for convenience, 
we first define a metric $\textbf{Z}_{d}^{\mathrm{SUM}}$ based on the weighted sum of the eigenscore of each sector $b$ and the \acp{RSRP} of each \acp{CCUAV} $d$,
as follows
\begin{gather}\label{eq:SUM_Metric}
    \tag{M1}
    \textbf{Z}_{d}^{\mathrm{SUM}} = \alpha \frac{ \mathrm{\textbf{ES}}_\mathrm{} - \mathrm{ES}_\mathrm{}^\mathrm{max}  }{  \mathrm{ES}_\mathrm{}^\mathrm{max} - \mathrm{ES}_\mathrm{}^\mathrm{min} } + \\
    + \left( 1-\alpha \right) \frac{ \mathrm{\textbf{RSRP}}_{d} - \mathrm{RSRP}_\mathrm{}^\mathrm{max}  }{  \mathrm{RSRP}_\mathrm{}^\mathrm{max} - \mathrm{RSRP}_\mathrm{}^\mathrm{min} }, \nonumber
\end{gather}
where 
$\alpha$ is a weighting factor,
$\textbf{ES}_\mathrm{} = \left[ \mathrm{ES}_1, \ldots, \mathrm{ES}_{N_\mathrm{MS}}\right]$
is the vector containing the eigenscores broadcasted by each sector $b$, 
$\textbf{RSRP}_d = \left[ \mathrm{RSRP}_1, \ldots, \mathrm{RSRP}_{N_\mathrm{MS}}\right]$
is the vector containing the \acp{RSRP} of \ac{CCUAV} $d$ with respect to each sector $b$,
$\mathrm{ES}_\mathrm{}^\mathrm{max} = \max\left( \mathrm{\textbf{ES}}_\mathrm{} \right)$
and $\mathrm{ES}_\mathrm{}^\mathrm{min} = \min\left( \mathrm{\textbf{ES}}_\mathrm{} \right)$.
%
Intuitively, 
the objective of this metric is to consider the eigenscore as a power offset to the traditional \ac{RSRP}-based metric during the cell selection process.
This would facilitate the computations at the \ac{CCUAV},
as the eigenscore, as mentioned earlier, can be broadcast by each sector in their control channels,
e.g., \ac{PBCH}.

The second defined metric $\textbf{Z}_d^{\mathrm{CAP}}$ based on the proposed eigenscore is inspired by the Shannon–Hartley channel capacity theorem, 
and has been formulated as follow
\begin{gather}\label{eq:CAP_Metric}
\small
    \tag{M2}
    \textbf{Z}_{d}^{\mathrm{CAP}} =
    \textbf{ES}_\mathrm{}  \log_2\left( 1 + \mathrm{\textbf{SNR}}_d  \right) \sim \\
    \sim \textbf{ES}_\mathrm{}  \log_2\left( 1 + \mathrm{\textbf{RSRP}}_d  \right). \nonumber
\end{gather} 
This metric attempts to assess the achievable capacity of \ac{CCUAV} $d$ in a noise-limited regime, 
taking into account the multiplexing capabilities of the sector on the route and the signal strength measured by the \ac{CCUAV}.
The computation of this metric is more involved than the previous one,
but allows to capture the linear and logarithmic relationship between spatial multiplexing and signal strength in terms of capacity.
No normalisation is needed. 


\section{Evaluation and Discussion}\label{sec:Results}
In this section, 
we evaluate the performance achieved by our proposed metrics when adopted to drive the cell selection process of \acp{CCUAV}. 
We adopt the scenario presented in Section \ref{sec:SystemModel}.
Note that $N_{\mathrm{ccuav}}=5$ \acp{CCUAV} are deployed on each aerial highway. 
The values of other parameters adopted in the numerical analysis are reported in Table~\ref{table:evaluation_paramerters}.
\begin{table}[!t]
\centering
\caption{Summary of the parameters used.\label{table:evaluation_paramerters}}
\begin{tabular}{|c c| c c |} 
 \hline
 \textbf{Param} & \textbf{Value} & \textbf{Param} & \textbf{Value} \\ [0.5ex] 
 \hline\hline
 A & 0.22~Km$^2$&      
 $N_{\rm MS}$ & 3\\        
 $M$ & 64 &
 $M_h$, $M_v$ & 4,4\\
 $\lambda_p$ & 8.57~cm&
$f_c$ & 3.5~GHz \\ 
 $N_g$ & 4 &
$L_r$ & 400~m \\
 
 $R$ & 18 &
 $N_w$ & 400 \\
 $\Delta_\phi$ & 10$^\circ$ &
$h_a$ & 1.5~m\\
  $d_\mathrm{ccuav}$ & 50~m & 
$h_a$ &  100~m\\
 $N_\mathrm{ccuav}$ & $\left\{1,\ldots, 7 \right\}$&
 $N_\mathrm{Drop}$ & 1000 \\
 $N_\mathrm{PRB}$ & 100&
  $P_{T_x}^\mathrm{Tot}$ & 46~dBm\\
 $P_{T_x}$ & 26~dBm&
 $K$ & 14.22 ~dB \\
 $\lambda_{Th}$ & 0.10&  
$\alpha$ & 0.50 \\
 \hline
\end{tabular}
\vspace{-2.5em}
\end{table}
Note that 
due to space constraints, 
we discuss the results obtained over a single aerial route,
that with $\Delta_\phi=90^\circ$.
However, similar results were obtained for all routes.

\subsection{Cell Selection Rate}\label{subsec:AssociationProbability}

In this section, 
we analyse the impact of the selected cell selection metric on the cell selection rate,
defined as the probability that an arbitrary \ac{CCUAV} selects a given sector as a serving one.

\begin{figure}[!t]
    \centering
    \subfigure[Cell selection rates using RSRP metric.]{
    \includegraphics[width=0.43\textwidth]{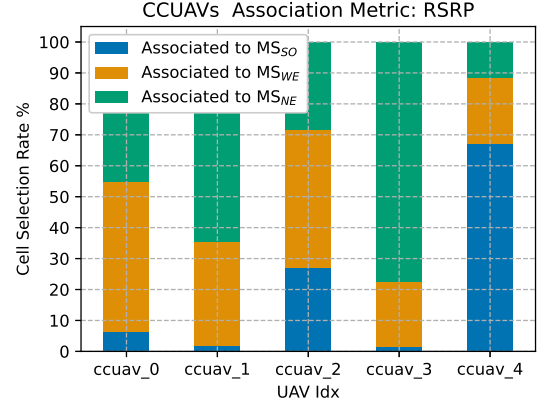}
    \label{subfig:associationProbRSRP}
    }
    \subfigure[Cell selection rates with summation based metric eq.~\eqref{eq:SUM_Metric}.]{
    \includegraphics[width=0.43\textwidth]{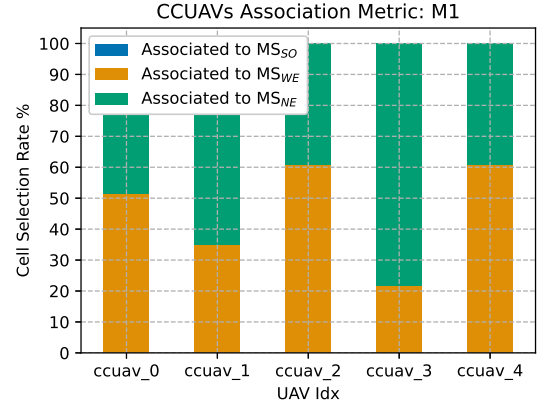}
    \label{subfig:associationProbM1}
    }
    \subfigure[Cell selection rates with capacity based metric eq.~\eqref{eq:CAP_Metric}.]{
    \includegraphics[width=0.43\textwidth]{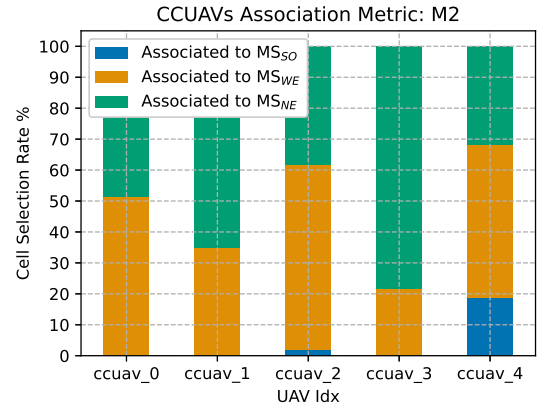}
    \label{subfig:associationProbM2}
    }
\caption{Cell selection rates of 5 CCUAVs on the vertical aerial highway (i.e., route rotated by $90^\circ$). \label{fig:AssociationProbability}}
\vspace{-1.5em}
\end{figure}

Figure~\ref{fig:AssociationProbability} shows the cell selection rate of the 5 deployed \acp{CCUAV} 
when considering three metrics: \ref{eq:SUM_Metric}, \ref{eq:CAP_Metric}, and \ac{RSRP}.
The first two metrics are the proposed ones, 
and the last one, noted as \ac{RSRP}, is the metric typically used in traditional networks,
which is used here as a benchmark.

As shown in Figure~\ref{subfig:associationProbRSRP},
when using the \ac{RSRP} metric,
some of the \acp{CCUAV} tend to associate to the south sector $\mathrm{MS}_{\mathrm{SO}}$, 
as it provides the strongest \ac{RSRP}.
For instance, 
$\mathrm{ccuav}_4$ connects $67.10\%$ of the time to the south sector $\mathrm{MS}_{\mathrm{SO}}$. 
When considering the other two studied metrics \ref{eq:SUM_Metric} and \ref{eq:CAP_Metric}, 
such rates dramatically change to $0.00\%$ and $18.60\%$, respectively.
Even if the south sector, $\mathrm{MS}_{\mathrm{SO}}$, may provide the largest received power,
the new eigenscore-based metrics lead to associations with sectors possessing better spatial resolution capabilities, 
i.e., $\mathrm{MS}_{\mathrm{WE}}$ and $\mathrm{MS}_{\mathrm{NE}}$,
as their \ac{mMIMO} panels have a better geometry with respect to the route,
As shown in Figure~\ref{fig:EigenvaluesEigenscore_evolution}, 
for the selected route, 
note that the south sector $\mathrm{MS}_{\mathrm{SO}}$ has the worst eigenscore, 
meaning that it cannot discern as many \acp{AoA} on the route as the other two sectors $\mathrm{MS}_{\mathrm{WE}}$ and $\mathrm{MS}_{\mathrm{NE}}$. 
Thus, associating more \acp{CCUAV} with the former sector leads to reduced spatial resolution, 
and ultimately reduced performance, 
as demonstrated in the following section.
%
%
%
%

It is worth highlighting that the results of this study also suggest that the two proposed metrics can be leveraged to reduce the number of candidate serving sectors for \acp{CCUAV}, 
resulting in fewer handovers and improved network stability. 
This can help to reduce overhead and handover failures, 
and thus enhance performance of the network.


\subsection{SINR Performance}

In the following, 
we analyse the \ac{UE} \ac{SINR} performance to further highlight the benefits of the proposed metrics when adopted to drive the \ac{CCUAV} cell selection process.

\begin{figure}[!t]
\vspace{-1em}
    \centering
    \includegraphics[width=0.49\textwidth]{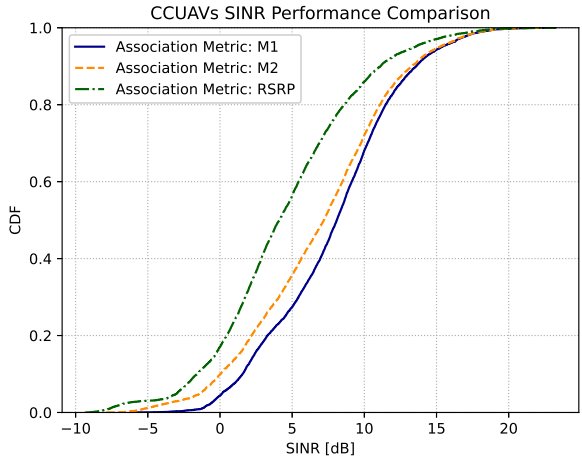}
    \caption{SINR comparison of CCUAVs placed on the vertical route.}
    \label{fig:SINR_dist_Vertical_route}
\vspace{-1.5em}
\end{figure}

Figure~\ref{fig:SINR_dist_Vertical_route} shows the \ac{SINR} distribution of the 5 deployed \acp{CCUAV} 
when considering the three metrics discussed earlier: \ref{eq:SUM_Metric}, \ref{eq:CAP_Metric}, and \ac{RSRP}.

The results show that when the proposed eigenscore-based metrics are adopted for driving the cell selection process,
an important increase in both average and 5\%-tile \acp{SINR} is achieved.
Specifically,
metric \ref{eq:SUM_Metric} achieves a gain of $3.30$\,dB and $3.13$\,dB with respect to the \ac{RSRP} metric at the average and 5\%-tile \acp{SINR}, respectively,
while the respective gains of metric \ref{eq:CAP_Metric} are $2.36$\,dB and $1.66$\,dB.
A summary of the \ac{CCUAV} \acp{SINR} and gains can be found in Table~\ref{table:SINR_dist_Results}.
\begin{table}
\vspace{1em}
\scriptsize
\centering
\caption{Summary of SINR results for different association metric.\label{table:SINR_dist_Results}}
\vspace{-0.5em}
\begin{tabular}{|c |c c c |} 
\hline 
  & \multicolumn{3}{|c|}{\textbf{Association Metric}} \\
 \textbf{ } & \textbf{M1} & \textbf{M2} & \textbf{RSRP}   \\
 \hline
 \hline
 \textbf{Aerial 5\%-tile SINR [dB]}& 0.16 & -1.31 & -2.97   \\ 
 \textbf{Gain 5\%-tile to RSRP [dB]}& 3.13 & 1.66 & --    \\ 
 \hline
 \textbf{Aerial mean SINR [dB]}& 7.79 & 6.85 & 4.49  \\ 
\textbf{ Gain to RSRP [dB]}& 3.30 & 2.36 & --    \\ 

\hline 
\end{tabular}

\vspace{-2em}
\end{table}

\begin{figure}[!t]
    \centering
    \includegraphics[width=0.49\textwidth]{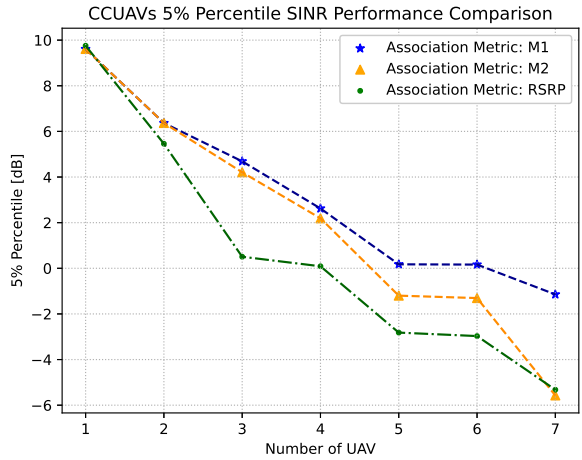}
    \caption{5\%-tile SINR versus the number of CCUAVs.}
    \label{fig:5Percentile_VS_NumberofCCUAVs}
\vspace{-0.5em}
\end{figure}

To further analyse the benefits of such metrics to support the reliable connectivity of multiple closely located \acp{CCUAV}, 
Figure~\ref{fig:5Percentile_VS_NumberofCCUAVs} shows how the 5\%-tile \ac{SINR} evolves when more and more \acp{CCUAV} fly on the same aerial highway 
(up to 7 \acp{CCUAV} with an inter-\ac{CCUAV} distance of $d_\mathrm{ccuav}=50$~m).
The results show that using the proposed metrics results in a significantly improved 5\%-tile SINR when compared to the \ac{RSRP} metric. 
Metrics~\ref{eq:CAP_Metric} and ~\ref{eq:SUM_Metric} achieve a maximum gain of $3.70$\,dB and $4.18$\,dB  for the case with 3 \acp{CCUAV} and 7 \acp{CCUAV}, respectively.  
A summary of the \ac{CCUAV} 5\%-tile SINR and respective gains can be found in Table~\ref{table:5Perc_Results}.

In summary, 
the improved performance when using the proposed metrics can be attributed to two factors. 
Firstly, 
the eigenscore enables the identification of cells that are more capable of effectively resolving \acp{AoA}, 
resulting in a higher multiplexing gain with \ac{mMIMO}.
Secondly, narrowing the pool of serving cell candidates reduces inter-cell interference. 

\begin{table}[!t]
\scriptsize
\centering
\caption{Summary of 5\%-tile SINR results for different association metric and number of CCUAVs.\label{table:5Perc_Results}}
\vspace{-0.5em}
\begin{tabular}{|c |c c c c c c c|} 
\hline 
  & \multicolumn{7}{|c|}{\textbf{Number of CCUAVs}} \\
\textbf{Metric}                         & \textbf{1} & \textbf{2} & \textbf{3} & \textbf{4} & \textbf{5} & \textbf{6} & \textbf{7} \\ \hline \hline
\textbf{M1 [dB]}                       & 9.61  & 6.37 & 4.69 & 2.62 & 0.17  & 0.16  & -1.14 \\
\textbf{M2 [dB]}                       & 9.61 & 6.37  & 4.21 & 2.19 & -1.20 & -1.31 & -5.57 \\
\textbf{RSRP [dB]}                     & 9.76  & 5.46 & 0.09 & 0.09 & -2.82 & -2.97 & -5.32 \\
\textbf{Gain M1-RSRP [dB]}      & -0.14 & 0.91 & 2.53 & 2.53 & 2.99  & 3.13  & 4.18 \\
\textbf{Gain M2-RSRP [dB]}     & -0.14 & 0.91 & 2.10 & 2.10 & 1.62  &  1.66 & -0.25 \\

\hline 
\end{tabular}
\vspace{-2.5em}
\end{table}

The results show that using \ac{RSRP} as the sole metric for choosing the serving cell among multiple suitable candidates is not the optimal solution. 
Integrating our proposed eigenscore into the cell association metric allows for better and fairer SINR performance in \ac{mMIMO}-based networks. 
In our results, 
using metric~\ref{eq:SUM_Metric} instead of~\ref{eq:CAP_Metric} yields better results. 
This suggests to network operators,
which plan to integrate aerial highway systems into their network, 
that they should incorporate such an eigenscore as an offset value to drive their cell selection and potentially handover processes.
Further studies are needed. 

\section{Conclusion}\label{sec:Conclusion}

In this paper, we have proposed a novel metric to drive the
cell selection process of \acp{UAV} on  aerial highways supported by a terrestrial \ac{mMIMO} network.
Our results showed that integrating our proposed metric, 
which is capable of capturing information on the spatial diversity between each aerial highway and sector, allows recognising and then associating with serving cells that provide a better and fairer \ac{SINR} performance,
especially when the number of flying \acp{CCUAV} increases. 
Future work will enhance our considerations,
by extending the metric also to embrace additional features such as information on the ground traffic condition. 


\bibliographystyle{IEEEtran}
\bibliography{journalAbbreviations, bibl}

\begin{acronym}[AAAAAAAAA]
    \acro{AoA}{angle of arrival}
    \acro{UAV}{unmanned aerial vehicle}
    \acro{CCUAV}{cellular connected unmanned aerial vehicle}
    \acro{D2D}{device to device}
    \acro{gUE}{ground user equipment}
    \acro{MIMO}{multiple-input multiple-output}
    \acro{mMIMO}{massive multiple-input multiple-output}
    \acro{PRB}{physical resource block}
    \acro{RSRP}{reference signal received power}
    \acro{RSS}{received signal strength}
    \acro{mmWave}{millimetre wave}
    \acro{eICIC}{enhanced inter-cell interference coordination}
    \acro{SINR}{signal-to-interference-plus-noise ratio}
    \acro{UAM}{urban air mobility}
    \acro{QoS}{quality of services}
    \acro{UE}{user equipment}
    \acro{LoS}{line of sight}
    \acro{BVLoS}{beyond visual line of sight}
    \acro{ZF}{zero forcing}
    \acro{CSI}{channel state information}
    \acro{3GPP}{3rd Generation Partnership Project}
    \acro{SVD}{single value decomposition}
    \acro{PBCH}{physical broadcast channel}
\end{acronym}
\end{document}